\documentclass[letter,11pt]{article}

\usepackage{amsmath,amssymb, amsthm}
\usepackage{complexity}
\usepackage{float}
\usepackage{algorithm}
\floatname{algorithm}{Protocol}
\usepackage{multirow}
\usepackage{graphicx}
\usepackage{amssymb,amsmath}
\usepackage[margin=1in]{geometry}
\usepackage[normalem]{ulem}

\newcommand{\graphstate} {brickwork state }
\newcounter{linenum}

\newcommand{\ket}[1]{\left| #1 \right\rangle}

\newcommand{\AliceSetInline}  { \{ 1/\sqrt{2} \left(\ket{0} + e^{i \theta} \ket{1} \right)
\mid \theta = 0, \pi/4, 2\pi/4, \ldots, 7\pi/4 \} }

\def\ost{\frac1{\sqrt2}}

\def\ost{\frac1{\sqrt2}}
\def\Cx#1{\cx{#1}{}}
\def\Cz#1{\cz{#1}{}}
\def\Cp#1#2{Z_{#1}(#2)}
\def\cz#1#2{Z_{#1}^{#2}}
\def\cx#1#2{X_{#1}^{#2}}
\def\cp#1#2#3{Z_{#1}^{#2}(#3)}

\def\m#1#2{{M}_{#2}^{#1}}
\def\M#1#2{{M}_{#2}^{#1}}
\def\et#1#2{E_{#1#2}}

\newtheorem{theorem}{Theorem}

\newtheorem{defn}{Definition}

\title{Universal Blind Quantum Computation}
\author{%
Anne Broadbent$^{1}$,
Joseph Fitzsimons$^{1, 2}$, %
Elham Kashefi $^{3}$
 \footnote{Email: \tt {albroadb@iqc.ca}, {joe.fitzsimons@materials.ox.ac.uk}, {ekashefi@inf.ed.ac.uk}}
}
\date{}

\begin{document}
\footnotetext[1]{Institute for Quantum Computing, University of Waterloo, Waterloo, Ontario, Canada.}
\footnotetext[2]{Materials Department, University of Oxford, Oxford, United Kingdom.}
\footnotetext[3]{School of Informatics, University of Edinburgh, Edinburgh, Scotland, United Kingdom.}
\maketitle

\thispagestyle{empty}
\begin{center}
\textbf{Abstract}
\end{center}
\noindent
\begin{quote}

We present a protocol which allows a client to have a server carry out a quantum computation for her such that the client's inputs, outputs and computation remain perfectly private, and where she does not require any quantum computational power or memory. The client only needs to be able to prepare single qubits randomly chosen from a finite set and send them to the server, who has the balance of the required quantum computational resources.  Our protocol is interactive: after the initial preparation of quantum states, the client and server  use two-way classical communication which enables the client to drive the computation, giving single-qubit measurement instructions to  the server, depending on previous measurement outcomes. Our protocol works for inputs and outputs that are either classical or quantum. We give an authentication protocol that allows the client to detect an interfering server; our scheme can also be made fault-tolerant.

We also generalize our result to the setting of a purely classical client who communicates classically with two non-communicating entangled servers, in order to perform a blind quantum computation.  By incorporating the authentication protocol, we show that any problem in $\BQP$ has an entangled two-prover interactive proof with a purely classical verifier.

Our protocol is the first universal scheme which detects a cheating server, as well as the first protocol which does not require any quantum computation whatsoever on the client's side.  The novelty of our approach is in using the unique features of measurement-based quantum computing which allows us to clearly distinguish between the quantum and classical aspects of a quantum computation.

\end{quote}
\newpage
\setcounter{page}{1}
\section{Introduction}
\label{sec:introduction}
 \pagestyle{plain}

When the technology to build quantum computers becomes
available, it is  likely that it will only be accessible to a
handful of centers around the world. Much like today's rental system
of supercomputers, users will probably be granted access to the
computers in a limited way. How will a
user interface with such a quantum computer?  Here, we consider the scenario where a user is
unwilling to reveal the computation that the remote computer is to perform,  but still wishes
to exploit this quantum resource.
More precisely, we give a protocol that allows a client Alice (who does not
have any quantum computational resources or quantum memory) to
interact with a server Bob (who has a quantum computer) in order for Alice to
obtain the outcome of her target computation such that privacy is
preserved. This means that Bob learns nothing about Alice's
inputs, outputs, or desired computation. The privacy is perfect,
does not rely on any computational assumptions, and holds no matter
what actions a cheating Bob undertakes. Alice only needs to be able to prepare single qubits randomly chosen from a finite set and send them to the server, who has the balance of the required quantum computational resources. After this initial preparation, Alice and Bob use  two-way classical communication which enables Alice to drive the computation by giving single-qubit measurement instructions to  Bob, depending on previous measurement outcomes.
Note that if Alice wanted to compute the solution to  a classical problem in
$\NP$, she could efficiently verify the outcome. An interfering Bob is not so obviously detected in other cases. We give an authentication technique which performs this detection. In order to make the protocol closer to practical implementations, we show how our scheme can be made fault-tolerant. Along the way, we also give a new universal family of graph states, the \emph{brickwork states} which, unlike the \emph{cluster states}, only require $(X,Y)$-plane measurements.

Our protocol can be used for any quantum circuit and also works for quantum inputs or outputs. We now give some applications.
\begin{itemize}
\item
\label{item:classical-classical-NP} \emph{Factoring}. Factoring is a
prime application of our protocol: by
implementing Shor's factoring algorithm~\cite{Shor} as a blind
quantum computation, Alice can use Bob to help her factor a product
of large primes which is associated with an RSA public key~\cite{RSA}.
Thanks to the properties of our protocol, Bob will not only be
unable to determine Alice's input, but will be completely oblivious
to the fact that he is helping her factor.

\item
\label{item:classical-classical-BQP} \emph{$\BQP$-complete problem.} Our protocol could be used to help Alice solve a
$\BQP$-complete problem, for instance
approximating the Jones polynomial~\cite{AVJ06}.  There is no known classical method to
efficiently verify the solution; this motivates the need for
authentication of Bob's computation,  even in the  case that the output is classical.

\item  \emph{Processing quantum information.} Alice may wish to use Bob as a remote device to manipulate quantum information. Consider the case where Alice is participating in a quantum protocol such as a quantum interactive proof. She can use our protocol to  prepare a quantum state, to perform a measurement on a quantum system, or to process quantum inputs into quantum outputs.
\end{itemize}
Our results also have direct applications to the domain of the complexity theory.
\begin{itemize}

\item \emph{Quantum prover interactive proofs.} Our scheme can be used to  accomplish an interactive proof for any language in $\BQP$, with a quantum prover and a nearly-classical verifier. The first interactive proof given explicitly in this scenario was proposed by Aharonov, Ben-Or and  Eban~\cite{Dorit} after a pre-print of our paper appeared.  In our early paper we did not consider the interactive proof scenario, however our protocol implicitly proposed an interactive proof for any language in $\BQP$ with a quantum prover and where the verifier requires the power to generate random qubits chosen from a fixed set, whereas the scheme in~\cite{Dorit}  requires a verifier with significantly more quantum power.

\item  \emph{Multi-prover interactive proofs.} Our protocol can be adapted to provide a two-prover interactive proof for any problem in $\BQP$ with a purely classical verifier. The modification requires that the provers share entanglement but otherwise be unable to communicate. Guided by the verifier, the first prover measures his part of the entanglement in order to create a shared resource between the verifier and the second prover. The remainder of the interaction involves the verifier and the second prover who essentially run our main protocol.

\end{itemize}

\subsection{Related work}

In the classical world, Feigenbaum introduced the notion of
\emph{computing with encrypted data}~\cite{Feigen86}, according to which a function~$f$ is
\emph{encryptable} if Alice can easily transform an instance~$x$
into instance $x'$, obtain $f(x')$ from Bob and efficiently compute
$f(x)$ from $f(x')$, in such a way that Bob cannot infer $x$ from
$x'$.
Following this, Abadi, Feigenbaum and Kilian~\cite{AFK} gave an impossibility result: no $\NP$-hard
function can be computed with encrypted data (even probabilistically and with polynomial interaction), unless the
polynomial hierarchy collapses at the third level.

Ignoring the blindness requirement of our protocol yields an interactive proof with a $\BQP$ prover and a nearly-classical verifier.  As mentioned, this scenario was first proposed in the work of \cite{Dorit}, using very different techniques based on authentication schemes.
Their protocol can be also used for blind quantum computation. However, their scheme requires that Alice have quantum computational resources and memory to act on a constant-sized register. A related classical protocol for the scenario involving a~$\P$ prover and a nearly-linear time verifier was given in~\cite{Muggles}.

Returning to the cryptographic scenario, still in the model where the function is classical and public,
Arrighi and Salvail~\cite{AS06} gave an approach using
quantum resources. The idea of their protocol is that Alice gives
Bob multiple quantum inputs, most of which are \emph{decoys}. Bob applies
the target function on all inputs, and then Alice verifies his
behaviour on the decoys. There are two important points to make here.
First, the protocol only works for a restricted set of classical
functions called \emph{random verifiable}: it must be possible for
Alice to efficiently generate random input-output pairs.
Second, the protocol does not prevent Bob
from learning Alice's private input; it provides only \emph{cheat
sensitivity}.

The case of a blind \emph{quantum} computation was first considered by  Childs~\cite{Childs05}
based on the idea of encrypting input qubits  with a quantum
one-time pad~\cite{AMTW00,BR00}. At each step, Alice sends the
encrypted qubits to Bob, who applies a known quantum gate  (some
gates requiring further interaction with Alice). Bob returns the
quantum state, which Alice decrypts using her key.
Cycling through a fixed set of universal gates ensures that Bob
learns nothing about the circuit. The protocol requires
fault-tolerant quantum memory  and the ability to apply local Pauli operators at each step, and does not provide any
method for the detection of malicious errors.

\subsection{Contributions and Techniques}
We present the first protocol for universal blind quantum
computation where Alice has no quantum memory. Our protocol works
for \emph{any} quantum circuit and assumes Alice has
a classical computer, augmented with the power to prepare single
qubits randomly chosen in~$$\AliceSetInline \,.$$ The required quantum and classical communication between Alice and Bob is linear in the size of Alice's desired quantum circuit.

Interestingly, it is
sufficient for our purposes to restrict Alice's classical
computation to modulo~8 arithmetic! Similar observations in a
non-cryptographic context have been made in~\cite{AB08}.
Except for an unavoidable leakage of the size of Alice's data \cite{AFK}, Alice's privacy is perfect.
We provide an authentication technique to detect an interfering Bob  with overwhelming probability;
this is optimal since there is always an exponentially small probability that
Bob can guess a path that will make Alice accept.
 We also show how the protocol can be made fault-tolerant.
 Furthermore, we extend our result to the domain of interactive proof systems: 
   we prove that any problem in $\BQP$ has an interactive proof system with two entangled provers and a purely classical verifier.

All previous protocols for blind quantum computation require
technology for Alice that is today unavailable: Arrighi and
Salvail's protocol  requires multi-qubit preparations and
measurements,  Childs' protocol requires
fault-tolerant quantum memory  and the ability to apply local Pauli operators at each step, while  Aharonov, Ben-Or and Eban's protocol requires a constant-sized quantum computer with memory.
In sharp contrast to this, from
Alice's point of view, our protocol can be implemented with physical
systems that are already available and well-developed. The required apparatus can be achieved by making only minor modifications to equipment used in the BB84 key exchange
protocol~\cite{BB84}. Single nitrogen vacancy centers in
diamond, for example, offer the necessary functionality and can be
used even at room temperature, removing the necessity for cumbersome
equipment such as cryostats~\cite{GOD+06}.

Our protocol is described in terms of the measurement-based model for quantum computation (MBQC) \cite{RB01,RBB03}.
While the computational power of this model is the same as in the
quantum circuit model~\cite{D89} (and our protocol could be completely recast into this model), it has proven to be conceptually
enlightening to reason about the distributed task of blind quantum
computation using this approach.
The novelty of our approach is in using the unique feature of MBQC that separates the classical and quantum parts of a computation, leading to a generic scheme for blind computation of any circuit without requiring any quantum memory for Alice. This is fundamentally different from previously known classical or quantum schemes.
Our protocol can be viewed as a distributed version of an MBQC computation (where Alice prepares the individual qubits, Bob does the entanglement and measurements, and Alice computes the classical feedforward mechanism), on top of which randomness is added in order to obscure the computation from Bob's point of view.
The family of graph states called \emph{cluster
states}~\cite{RB01} is universal for MBQC (\emph{graph states} are initial entangled states required for the
computation in MBQC). However, the method that
allows arbitrary computation on the cluster state consists in first
tailoring the cluster state to the specific computation by
performing some computational basis measurements. If we were to
use this principle for blind quantum computing, Alice would have to
reveal information about the structure of the underlying graph
state. We introduce a new  family of states called the
\emph{brickwork states} (Figure \ref{fig:cluster})
which are universal for $X-Y$ plane measurements and thus do not require the initial  computational basis measurements. Other universal graph states for  that do not require initial computational basis measurements have appeared in~\cite{CLN05}.

To the best of our knowledge, this is the first time that a new functionality has been achieved thanks to MBQC (other theoretical advances due to MBQC appear in~\cite{RHG06, MS08}). From a conceptual point of view, our contribution shows that MBQC has tremendous potential for the development of new  protocols, and maybe even of algorithms.

\subsection{Outline of Protocols}

The outline of the main protocol is as follows. Alice has in mind a
quantum computation given as a \emph{measurement pattern} on a  \emph{brickwork} state.
There are two stages:
\emph{preparation} and \emph{computation}. In the preparation stage,
Alice prepares single qubits  chosen randomly from   $\AliceSetInline$ and
sends them to Bob. After receiving all the qubits, Bob entangles
them according to the brickwork state. Note that this unavoidably
reveals upper bounds on the dimensions of Alice's underlying graph
state, that correspond to the length of the input and depth of the
computation. However, due to universality of the brickwork state, it does not reveal any
additional information on Alice's computation. The computation stage
involves interaction: for each layer of the brickwork state, for
each qubit, Alice sends a classical message to Bob to tell him in
which basis of the~$X-Y$ plane he should measure the qubit. Bob
performs the measurement and  communicates the outcome; Alice's
choice of angles in future rounds will depend on these values.
Importantly, Alice's quantum states and classical messages are
astutely chosen so that, no matter what Bob does, he cannot infer
anything about her measurement pattern. If Alice is computing a
classical function, the protocol finishes when all  qubits are
measured. If she is computing a quantum function, Bob returns to her
the final qubits. A modification of the protocol also allows Alice's
inputs to be quantum.

We give an authentication technique which
enables Alice to detect an interfering Bob with overwhelming
probability (strictly speaking, either Bob's interference is
corrected and he is not detected, or his interference is detected
with overwhelming probability). The authentication  requires that Alice encode her input into an error
correction code and choose an appropriate fault-tolerant implementation of her desired computation. She also uses some qubits as
\emph{traps}; they are prepared in the eigenstates of the
Pauli operators~$X$, $Y$ and~$Z$.

The remainder of the paper is structured as follows:
the main protocol is given in Section~\ref{sec:protocol}, where universality,
correctness and blindness are proven. Section~\ref{sec:extensions} discusses extensions to the case of
quantum inputs or outputs; authentication techniques that
are used to detect an interfering Bob and perform fault-tolerant computations are in Section~\ref{sec:auth-fault-tolerance},
while Section~\ref{sec:entangled-servers} presents the two-server protocol with a purely classical Alice. The reader unfamiliar with MBQC  is referred to a short introduction in Appendix~\ref{appendix:measurement calculus}. Appendix~\ref{appendix:universal} contains a universality proof of the brickwork states that is lengthy due to its figures.


\section{Main Protocol}
\label{sec:protocol}
\begin{figure} \centering
   \includegraphics[scale=0.8]{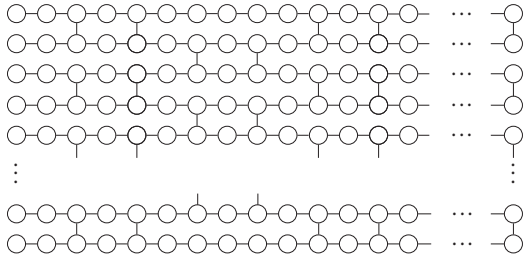}

\caption{{The  \emph{brickwork state, $\mathcal{G}_{n \times m}$}.
    Qubits $\ket{\psi_{x,y}}$ $(x=1, \ldots, n, y=1,\ldots,m)$ are arranged according to layer $x$ and row $y$,
     corresponding
to the vertices in the above graph, and are originally in
the~$\ket{+} = \frac{1}{\sqrt{2}}\ket{0}
+\frac{1}{\sqrt{2}}\ket{1}$ state.
    Controlled-$Z$ gates are then performed
     between qubits which are joined by an edge.
     }
    \label{fig:cluster}}
\end{figure}

Suppose Alice has in mind a unitary operator $U$ that is implemented
with a pattern  on a brickwork state $\mathcal{G}_{n
\times m}$ (Figure~\ref{fig:cluster}) with measurements given as multiples of~$\pi/4$.
This pattern could have been designed either directly in MBQC or from a circuit
construction.
 Each qubit $\ket{\psi_{x,y}} \in  \mathcal{G}_{n \times m}$ is
indexed by a \emph{column} $x \in \{1, \ldots ,n\}$ and a
\emph{row}~$y \in \{1, \ldots ,m\}$. Thus each qubit is assigned: a measurement
angle~$\phi_{x,y}$, a set of $X$-dependencies $D_{x,y} \subseteq [x-1]\times[m]$, and
a set of $Z$-dependencies $D'_{x,y} \subseteq [x-1]\times [m]$\,.
Here, we assume that the dependency sets~$X_{x,y}$ and~$Z_{x,y}$ are obtained via the
flow construction~\cite{Flow06}.
During the execution of
the pattern, the actual measurement angle $\phi'_{x,y}$ is a
modification of $\phi_{x,y}$ that depends on previous measurement
outcomes in the following way: let $s^X_{x,y} = \oplus_{i\in
D_{x,y}}{s_i}$ be the parity of all measurement outcomes for qubits
in $X_{x,y}$ and similarly, $s^Z_{x,y} = \oplus_{i\in D'_{x,y}}{s_i}$ be
the parity of all measurement outcomes for qubits in~$Z_{x,y}$. Then
$ \phi'_{x,y} = (-1)^{s^X_{x,y}} \phi_{x,y} + s^Z_{x,y} \pi$\,. \textbf{Protocol~\ref{prot:UBQC}} implements a blind quantum computation for~$U$.
Note that we assume that  Alice's input to the computation is built into $U$. In other words, Alice wishes to compute $U \ket{0}$, her input is classical and the first layers of~$U$ may depend on it.

\renewcommand{\labelenumi}{\textbf{\arabic{enumi}.}}
\renewcommand{\labelenumii}{\arabic{enumi}.\arabic{enumii}}
\renewcommand{\labelenumiii}{\arabic{enumi}.\arabic{enumii}.\arabic{enumiii}}

\begin{algorithm}
\caption{Universal Blind Quantum Computation}
 \label{prot:UBQC}

\begin{enumerate}

\item  \label{step:Alice-prep} \textbf{Alice's preparation} \\
For each column $x = 1, \ldots , n$ \\
\hspace*{\parindent}\hspace*{\parindent} For each row $y = 1,
\ldots , m$
\begin{enumerate}
\item \label{step:one-one}Alice prepares  $\ket{\psi_{x,y}}
 \in_R  \{ \ket{+_{\theta_{x,y}}}= \frac{1}{\sqrt{2}}(\ket{0} + e^{i\theta_{x,y}}\ket{1}) \mid \theta_{x,y} = 0,
\pi/4, \ldots, 7\pi/4 \}$ and sends the qubits to Bob.
\end{enumerate}

\item  \label{step:Bob-prep} \textbf{Bob's preparation} %
\begin{enumerate}
\item Bob creates an entangled state from all received qubits, according to their indices,
by applying \textsc{ctrl}-$Z$ gates between the qubits in order to
create a brickwork state $\mathcal{G}_{n \times m}$ (see Definition~\ref{defn:brick}).
\end{enumerate}

\item  \label{step:computation-interaction} \textbf{Interaction and measurement}
\\
For each column $x = 1, \ldots , n$ \\
\hspace*{\parindent}\hspace*{\parindent} For each row $y = 1,
\ldots , m$
\begin{enumerate}
\item \label{step:three-one} Alice computes $\phi'_{x,y}$
where $s^X_{0,y}=s^Z_{0,y}=0$.
\item Alice chooses $r_{x,y} \in_R \{0,1\}$ and computes $\delta_{x,y} =
\phi'_{x,y}  + \theta_{x,y} + \pi r_{x,y}$\,.
\item Alice  transmits $\delta_{x,y}$ to Bob. Bob
measures in the basis $\{ \ket{+_{\delta_{x,y}}},
\ket{-_{\delta_{x,y}}} \}$.
\item Bob transmits the result $s_{x,y} \in \{0,1\}$ to Alice.
\item \label{one-timepad}If $r_{x,y} = 1$ above, Alice flips $s_{x,y}$;
otherwise she does nothing.
\end{enumerate}
\end{enumerate}
\end{algorithm}

The universality of  \textbf{Protocol~\ref{prot:UBQC}} follows from the universality of brickwork state (defined below) for measurement-based quantum computing.

\begin{defn} \label{defn:brick} A brickwork state $\mathcal{G}_{n \times m}$, where $m \equiv 5 \pmod 8$, is an entangled state of~$n\times m$ qubits constructed as follows (see also Figure~\ref{fig:cluster}):
\begin{enumerate}
\item Prepare all qubits in state $\ket +$ and assign to each qubit an index $(i,j)$,  $i$ being a column ($i \in [n]$) and $j$ being  a row ($j \in [m]$).
\item For each row, apply the operator \textsc{ctrl}-$Z$ on qubits $(i,j)$ and $(i,j+1)$ where $1\leq j \leq m-1$.
\item For each column $j \equiv 3 \pmod 8$ and each odd row $i$, apply the operator \textsc{ctrl}-$Z$ on qubits $(i,j)$ and $(i+1,j)$ and also on qubits $(i,j+2)$ and $(i+1,j+2)$.
\item For each column $j \equiv 7 \pmod 8$ and each even row $i$, apply the operator \textsc{ctrl}-$Z$ on qubits $(i,j)$ and $(i+1,j)$ and also on qubits $(i,j+2)$ and $(i+1,j+2)$.
\end{enumerate}
\end{defn}

The proof of the following theorem is relegated to Appendix~\ref{appendix:universal}. 

\begin{theorem}[Universality]
\label{thm:universal} The \graphstate $\mathcal{G}_{n \times m}$ is
universal for quantum computation. Furthermore, we only require
single-qubit measurements under the angles $\{0, \pm \pi/4,  \pm \pi/2\}$,
and measurements can be done layer-by-layer.
\end{theorem}

In this work, we only consider approximate universality. This  allows us to restrict the angles of preparation and measurement to a finite set and hence simplify the description of the protocol. However one can easily extend our protocol to achieve exact universality as well, provided Alice can communicate real numbers to Bob.

\emph{Correctness} refers to the fact that the outcome of the protocol is the same as the outcome if Alice had run the pattern herself. The fact that Protocol~\ref{prot:UBQC} correctly computes $U\ket{0}$  follows from the commutativity of Alice's rotations and Bob's measurements in the rotated bases. This is formalized~below.

\begin{theorem}[Correctness]
\label{thm:Correctness}Assume Alice and Bob follow the steps of
\textbf{Protocol~\ref{prot:UBQC}}. Then the outcome is correct.
\end{theorem}
\begin{proof} Firstly, since  \textsc{ctrl}-$Z$ commutes with $Z$-rotations,
steps~\ref{step:Alice-prep} and
\ref{step:Bob-prep} do not change the underlying graph
state; only the phase of each qubit is locally changed, and it is as
if Bob had done the $Z$-rotation after the  \textsc{ctrl}-$Z$.
Secondly, since a measurement in the $\ket{+_\phi}, \ket{-_\phi}$
basis on a state $\ket{\psi}$ is the same as a measurement in the $\ket{+_{\phi  +\theta}},
\ket{-_{\phi  +\theta}}$ basis on $Z(\theta) \ket{\psi}$, and
since $\delta = \phi' + \theta + \pi r$, if $r=0$, Bob's measurement
has the same effect as Alice's target measurement; if $r=1$,
all Alice needs to do is flip the outcome.
\end{proof}

We now define and prove the security of the protocol.  Intuitively, we wish to prove that whatever Bob chooses to do (including
arbitrary deviations from the protocol), his knowledge on Alice's quantum computation does not increase.  Note, however that Bob does learn the dimensions of the
brickwork state, giving an upper bound on the size of Alice's
computation. This is unavoidable: a simple adaptation of the proof of Theorem~2 from~\cite{AFK},
confirms this. We incorporate this notion of leakage in our definition of \emph{blindness}. A \emph{quantum delegated computation} protocol is a protocol by which Alice interacts quantumly with Bob in order to obtain the result of a computation, $U(x)$, where $X=(\tilde{U},x)$ is Alice's input with $\tilde{U}$ being a description of $U$.

\begin{defn} \label{defn:blind}Let \textsf{P} be a quantum delegated computation on input $X$ and let $L(X)$ be any function of the input. We say that a quantum delegated computation protocol is \emph{blind while leaking at most L(X)} if, on Alice's input X, for any fixed $Y=L(X)$, the following two hold when given $Y$:
\begin{enumerate}
\item \label{defn:blind-1} The distribution of the classical information obtained by Bob in \textsf{P} is independent of $X$.
\item Given the distribution of classical information described in~\ref{defn:blind-1}, the state of the quantum system obtained by Bob in \textsf{P} is fixed and independent of~$X$.
\end{enumerate}
\end{defn}

Definition~\ref{defn:blind} captures the intuitive notion that Bob's
view of the protocol should not depend on~$X$ (when given $Y$); since his view consists of classical and quantum information, this means that the distribution of the classical information should not depend on $X$ (given $Y$) and that for any fixed choice of the classical information, the state of the quantum system should be uniquely determined and not depend on $X$ (given~$Y$). We are now ready to state and prove our main theorem. Recall that in \textbf{Protocol~\ref{prot:UBQC}}, $(n,m)$ is the dimension of the brickwork state.

\begin{theorem}[Blindness]\label{thm:privacy}
\textbf{Protocol~\ref{prot:UBQC}} is blind while leaking at most $(n,m)$.
\end{theorem}
\begin{proof}
Let $(n,m)$ (the dimension of the brickwork state) be given. Note that the universality of the brickwork state guarantees
that Bob's creating of the graph state does not reveal anything on the underlying computation (except $n$ and $m$).

Alice's input consists of $$\phi = (\phi_{x,y} \mid  x \in  [n], y \in [m])$$ with the actual measurement angles $$\phi'= (\phi'_{x,y} \mid  x \in  [n], y \in [m])$$ being
a modification of $\phi$ that depends on previous measurement
outcomes. Let the classical information that Bob gets during the protocol be $$\delta = (\delta_{x,y} \mid x \in  [n], y \in [m])$$ and let $A$ be the quantum system initially sent from Alice to Bob.

To show independence of Bob's classical information, let $\theta'_{x,y} = \theta_{x,y} + \pi r_{x,y}$ (for a uniformly random chosen $\theta_{x,y}$) and  $\theta'= (\theta'_{x,y} \mid  x \in  [n], y \in [m])$. We have  $\delta = \phi' + \theta'$, with $\theta'$ being uniformly random (and independent of  $\phi$ and/or~$\phi'$), which implies the independence of  $\delta$ and~$\phi$.

As for Bob's quantum information, first fix an arbitrary choice of $\delta$.  Because  $r_{x,y}$ is uniformly  random,
for each qubit of $A$, one of the following two has occurred:

\begin{enumerate}
\item $r_{x,y} = 0$ so $\delta_{x,y} =\phi'_{x,y} + \theta'_{x,y}$ and          $\ket{\psi_{x,y}} =\frac{1}{\sqrt{2}}(\ket{0} + e^{i(\delta_{x,y} - \phi'_{x,y})}\ket{1}$.
\item  $r_{x,y} = 1$ so  $\delta_{x,y} =\phi'_{x,y} + \theta'_{x,y}  + \pi$ and $\ket{\psi_{x,y}} = \frac{1}{\sqrt{2}}(\ket{0} - e^{i(\delta_{x,y} - \phi'_{x,y})}\ket{1}$.
\end{enumerate}
Since $\delta$ is fixed, $\theta'$ depends on $\phi'$ (and thus on~$\phi$), but since $r_{x,y}$ is independent of everything else, without knowledge of $r_{x,y}$ (i.e.~taking the partial trace of the system over Alice's secret), $A$ consists of copies of the two-dimensional completely mixed state, which is fixed and independent of~$\phi$.
\end{proof}

There are two malicious scenarios that are covered by  Definition~\ref{defn:blind} and that we explicitly mention here. Suppose Bob has some prior knowledge, given as some  \emph{a priori} distribution on Alice's input~$X$.   Since Definition~\ref{defn:blind} applies to any distribution  of $X$,  we can simply apply it to the conditional distribution representing the distribution of $X$ given Bob's a priori knowledge; we conclude that Bob does not learn any information on $X$ beyond what he already knows, as well as what is leaked. The second scenario concerns a Bob whose goal it is to find Alice's output. Definition~\ref{defn:blind} forbids this: learning information on the output would imply learning information on Alice's input.

Note that the protocol does not allow Alice to reveal to Bob whether or not she accepts the result of the computation as this bit of information could be exploited by Bob to learn some information about the actual computation. In this scenario, \textbf{Protocol \ref{prot:UBQC-authentication}} can be used instead.

\section{Quantum Inputs and Outputs}
\label{sec:extensions}

We can slightly modify \textbf{Protocol~\ref{prot:UBQC}} to deal with both quantum inputs and outputs. In the former case, no extra channel resources are required, while the latter case requires
a quantum channel from Bob to Alice in order for him to return the
output qubits. Alice will also need to be able to apply $X$ and $Z$
Pauli operators in order to undo the quantum one-time pad.  Note that these protocols can be combined to obtain a protocol for quantum inputs and outputs.

\subsection{Quantum Inputs}

Consider the scenario where Alice's input  is  the form of~$m$ physical qubits and she has no efficient classical description of
the inputs to be able to incorporate it into \textbf{Protocol
\ref{prot:UBQC}}. In this case, she needs to be able to apply local Pauli-$X$ and Pauli-$Z$ operators to
implement a full one-time pad over the input qubits. The first layer of measurements are adapted to undo the
Pauli-X
operation if necessary.  By the quantum one-time pad, Theorem~\ref{thm:Correctness} and Theorem~\ref{thm:privacy}, this modified protocol, given in \textbf{Protocol~\ref{prot:UBQC-qinput}}  is still correct and
private.

Here we assume that Alice already has in her hands the quantum
inputs: unless she receives the inputs one-by-one, she requires for
this initial step some quantum memory. She also needs to be able to
apply the single-qubit gates as described above. Note that this is
only asking slightly more than Alice choosing between four single-qubit gates, which would be the minimum required in \emph{any} blind quantum
computation protocol with quantum inputs.

\begin{algorithm}
\caption{Universal Blind Quantum Computation with Quantum Inputs}
 \label{prot:UBQC-qinput}

\begin{enumerate}
\item  \label{step:Alice-input1} \textbf{Alice's input preparation} \\
For the input column ($x=0, y=1, \ldots, m$) corresponding to Alice's input
\begin{enumerate}
\item \label{step:input-rotation1}Alice applies  $Z_{0,y}(\theta_{0,y})$ for $\theta_{0,y} \in_R \{ 0, \pi/4, 2\pi/4, \ldots, 7\pi/4 \}$.
\item \label{step:input-onetimepad1} Alice chooses $i_{0,y} \in_R \{0,1\}$ and applies
$X^{i_{0,y}}_{0,y}$. She sends the qubits to Bob.
\end{enumerate}

\item  \label{step:Alice-prep-aux1} \textbf{Alice's auxiliary preparation} \\
For each column $x = 1, \ldots , n$ \\
\hspace*{\parindent} For each row $y = 1, \ldots , m$
\begin{enumerate}
\item \label{step:one-one1}Alice prepares  $\ket{\psi_{x,y}}
 \in_R  \{ \ket{+_{\theta_{x,y}}} \mid \theta_{x,y} = 0,
\pi/4, 2\pi/4, \ldots, 7\pi/4 \}$ and sends the qubits to Bob.
\end{enumerate}

\item  \label{step:Bob-prep1} \textbf{Bob's preparation} %
\begin{enumerate}
\item Bob creates an entangled state from all received qubits, according to their indices, by applying \textsc{ctrl}-$Z$ gates between the qubits in order to
create a \graphstate $\mathcal{G}_{(n+1) \times m}$.
\end{enumerate}

\item  \label{step:computation-interaction1} \textbf{Interaction and measurement}
\\
For each column $x = 0, \ldots , n$ \\
\hspace*{\parindent} For each row $y = 1, \ldots , m$
\begin{enumerate}
\item \label{step:three-one1} Alice computes $\phi'_{x,y}$ with the special case  $\phi'_{0,y} =
(-1)^{i_{0,y}}\phi_{0,y}$.
\item Alice chooses $r_{x,y} \in_R \{0,1\}$ and computes $\delta_{x,y} =
\phi'_{x,y}  + \theta_{x,y} + \pi r_{x,y}$\,.
\item Alice  transmits $\delta_{x,y}$ to Bob.
\item \label{step:three-three}Bob measures in the basis $\{ \ket{+_{\delta_{x,y}}},  \ket{-_{\delta_{x,y}}}
\}$.

\item Bob transmits the result $s_{x,y} \in \{0,1\}$ to Alice.
\item If $r_{x,y} = 1$ above, Alice flips $s_{x,y}$;
otherwise she does nothing.
\end{enumerate}

\end{enumerate}
\end{algorithm}

\subsection{Quantum Outputs}

Suppose Alice now requires a quantum output, for example in the case of blind quantum  state preparation. In this scenario, instead of measuring the last layer of  qubits, Bob returns it to Alice, who  performs the final layer
of Pauli corrections. The following theorem shows a privacy property on the quantum states that Bob manipulates.

\begin{theorem}
\label{thm-onetimepadd} At every step of \textbf{Protocol
\ref{prot:UBQC}}, Bob's quantum state is one-time padded.
\end{theorem}
\begin{proof}
During the execution of the protocol the value of $s^X$ and $s^Z$ are unknown to Bob since they have been one-time padded using the random key $r$ at each layer. Due to the flow construction~\cite{Flow06}, each qubit (starting at the third column) receives independent Pauli operators, which act as the full quantum one-time pad over Bob's state. Since
our initial state is~$\ket{+}$, and since the first layer performs a hidden $Z$-rotation, it follows that the qubits in the second layer are also completely encrypted during the computation.
\end{proof}
This result together with Theorems~\ref{thm:Correctness} and \ref{thm:privacy} proves the correctness and privacy of \textbf{Protocol~\ref{prot:UBQC-qoutput}} that deals with quantum outputs.

\begin{algorithm}
\caption{Universal Blind Quantum Computation with Quantum Outputs}
 \label{prot:UBQC-qoutput}

\begin{enumerate}

\item   \textbf{Alice's auxiliary preparation} \\
For each column $x = 1, \ldots , n-1$ \\
\hspace*{\parindent} For each row $y = 1, \ldots , m$
\begin{enumerate}
\item Alice prepares  $\ket{\psi_{x,y}}
 \in_R  \{ \ket{+_{\theta_{x,y}}} \mid \theta_{x,y} = 0,
\pi/4, 2\pi/4, \ldots, 7\pi/4 \}$ and sends the qubits to Bob.
\end{enumerate}

\item   \textbf{Alice's output preparation}
\begin{enumerate}
\item Alice prepares the last column of qubits $\ket{\psi_{n,y}} =
\ket{+}$ ($y = 1, \ldots, m$) and sends the qubits to Bob.
\end{enumerate}

\item  \textbf{Bob's preparation} %
\begin{enumerate}
\item Bob creates an entangled state from all received qubits, according to their indices, by applying \textsc{ctrl}-$Z$ gates between the qubits in order to
create a \graphstate $\mathcal{G}_{n \times m}$.
\end{enumerate}

\item  \label{step:computation-interaction2} \textbf{Interaction and measurement}
\\
For each column $x = 1, \ldots , n-1$ \\
\hspace*{\parindent} For each row
 $y = 1,
\ldots , m$
\begin{enumerate}
\item \label{step:three-one2} Alice computes $\phi'_{x,y}$ where $s^X_{0,y}=s^Z_{0,y}=0$ for the first column.
\item Alice chooses $r_{x,y} \in_R \{0,1\}$ and computes $\delta_{x,y} =
\phi'_{x,y}  + \theta_{x,y} + \pi r_{x,y}$\,.
\item Alice  transmits $\delta_{x,y}$ to Bob.
\item \label{step:three-three2}Bob measures in the basis $\{ \ket{+_{\delta_{x,y}}},  \ket{-_{\delta_{x,y}}}
\}$.

\item Bob transmits the result $s_{x,y} \in \{0,1\}$ to Alice.
\item If $r_{x,y} = 1$ above, Alice flips $s_{x,y}$;
otherwise she does nothing.
\end{enumerate}
\item \label{step:output2} \textbf{Output Correction}
\begin{enumerate}
\item Bob sends to Alice all qubits in the last layer.
\item Alice performs the final Pauli corrections $Z^{s^Z_{n,y}}X^{s^X_{n,y}}$.
\end{enumerate}

\end{enumerate}
\end{algorithm}

\section{Authentication and Fault-Tolerance}
\label{sec:auth-fault-tolerance}

We now focus on Alice's ability to detect if Bob is not cooperating.
There are two possible ways in which Bob can be uncooperative:  he can refuse to perform
the computation (this is immediately apparent to Alice), or he can actively
interfere with the computation, while pretending to follow the
protocol. It is this latter case that we focus on detecting.
Our authentication technique enables Alice to detect an interfering Bob with overwhelming
probability (strictly speaking, either Bob's interference is
corrected and he is not detected, or his interference is detected
with overwhelming probability).  Note that this is the best that we
can hope for since nothing prevents Bob from refusing to perform the
computation. Bob could also be lucky and guess a path that Alice
will accept. This happens with exponentially small probability,
hence our technique is optimal.

In the case that Alice's computation has a classical output and that she does not require fault-tolerance, a simple  protocol for blind quantum computing with authentication exists: execute \textbf{Protocol~\ref{prot:UBQC}}, on a modification of Alice's target circuit: she adds~$N$ randomly placed trap wires that are randomly in state~$\ket{0}$ or $\ket{1}$ ($N$ is the number of  qubits in the computation). If Bob interferes, either his interference has no effect on the classical output, or he will get caught with probability at least~$\frac{1}{2}$ (he gets caught if Alice finds that the output of at least one trap wire is incorrect).  The protocol is repeated $s$ times (the traps are randomly re-positioned each time); if Bob is not caught cheating, Alice accepts if all outputs are identical; otherwise she rejects. The probability of an incorrect output being accepted is at most $2^{-s}$.

Our contribution, \textbf{Protocol~\ref{prot:UBQC-authentication}}  is more general than this scheme since it works for quantum  outputs and is fault-tolerant. If the above scheme is used for quantum inputs, they must be given to Alice as multiple copies. Similarly (but more realistically), if \textbf{Protocol~\ref{prot:UBQC-authentication}} is to be used on quantum inputs, these must already be given to Alice in an encoded form as in \textbf{step~\ref{step:auth-encode}} of \textbf{Protocol~\ref{prot:UBQC-authentication}} (because Alice has no quantum computational power). In the case of a quantum output, it will be given to Alice in a known encoded form, which she can pass on to a third party for verification.

The theory of quantum error correction provides a natural mechanism for detecting
unintended changes to a computation, whereas the theory of
fault-tolerant computation provides a way to process information even using error-prone gates.
Unfortunately, error correction, even when combined
with fault-tolerant gate constructions is insufficient to detect
malicious tampering if the error correction code is known. As
evidenced by the quantum authentication protocol~\cite{BCGST02},
error correction encodings can, however, be adapted for this purpose.

\begin{algorithm}
\caption{Blind Quantum Computing with Authentication (classical input and output)
}\label{prot:UBQC-authentication}

\begin{enumerate}
\item Alice chooses $\mathbb{C}$, where
 $\mathbb{C}$ is some $n_\mathbb{C}$-qubit error-correcting code with
 distance~$d_\mathbb{C}$. The security parameter is $d_\mathbb{C}$.
 \item \label{step:auth-encode} In the circuit model, starting from circuit for $U$,  Alice converts target circuit to fault-tolerant circuit:
  \begin{enumerate}
	\item Use error-correcting code $\mathbb{C}$. The encoding appears in the initial layers of the circuit.
	\item Perform all gates and measurements fault-tolerantly.
	\item Some computational basis measurements are required for the fault-tolerant implementation (for verification of ancillae and non-transversal gates).
Each measurement is accomplished by making and measuring a  \emph{pseudo-copy} of the target qubit: a \textsc{ctrl}-$X$ is performed from the target to an ancilla qubit initially set to $\ket{0}$, which is then measured in the  $Z$-basis.
	\item Ancilla qubit wires are evenly spaced through the circuit.
	\item The ancillae are re-used. All ancillae are measured at the same time, at regular intervals, after each fault-tolerant gate (some outputs may be meaningless).
  \end{enumerate}

\item Within each encoded qubit, permute all wires,  keeping these permutations secret from Bob.
\item Within each encoded qubit, add  $3n_T$  randomly interspersed \emph{trap} wires, each trap being a random eigenstate of $X$, $Y$ or $Z$ ($n_T$ of each).
For security, we must have  $n_T \propto n_\mathbb{C}$; for  convenience, we choose  $n_T = n_\mathbb{C}$.
The trap qubit wire (at this point) does not interact with the rest of the circuit. The wire is initially $\ket{0}$, and then single-qubit gates are used to create the trap state. These single-qubit gates appear in the initial layers of the circuit.
\item Trap qubits are verified using the same ancillae as above: they are rotated into the computational basis, measured using the pseudo-copy technique above, and then returned to their initial basis.
\item Any fault-tolerant measurement is randomly interspersed with verification of $3n_T$ random trap wires. For this, identity gates are added as required.
\item For classical output, the trap wires are rotated as a last step, so that the following  measurement in the computational basis is used for a final verification.
\item \label{step:convert-to-MBQC} Convert the whole circuit above to a measurement-based computation on
 the brickwork state, with the addition of regular $Z$-basis measurements
 corresponding to the measurements on ancillae qubits above. Swap and identity gates are added as required, and trap qubits are left untouched.
 \item \label{auth:Blind-QC}Perform the blind quantum computation:
 \begin{enumerate}
\item Execute \textbf{Protocol~\ref{prot:UBQC}}, to which we add that Alice periodically instructs Bob to measure in~$Z$-basis as indicated above.
\item Alice uses the results of the trap qubit measurements to estimate the error rate; if it is below the threshold (see discussion in the main text), she accepts, otherwise she rejects.
\end{enumerate}

\end{enumerate}
\end{algorithm}

Our protocol proceeds along the following lines.
 Alice chooses
an $n_\mathbb{C}$-qubit error correction code~$\mathbb{C}$ with
distance $d_\mathbb{C}$. (The values of $n_\mathbb{C}$ and
$d_\mathbb{C}$ are taken as security parameters.) If the original
computation involves $N$ logical qubits, the authenticated version
 involves $N (n_\mathbb{C} + 3 n_T)$ (with $n_T= n_\mathbb{C}$), logical qubits: throughout the
computation, each logical qubit is encoded with~$\mathbb{C}$, while
the remaining $3 N n_T$ qubits are used as traps to detect an
interfering Bob. The trap qubits are prepared as a first step
of the computation in eigenstates of the
Pauli operators $X$, $Y$ and $Z$,  with an equal number of qubits in each state.

The protocol also involves fault-tolerant gates, for some of which it
is necessary to have Bob periodically measure qubits~\cite{ZCC2007a}. In order to
accomplish this, the blind computation protocol is extended by
allowing Alice to instruct Bob to measure specific qubits within the
brickwork state in the computational basis at regular intervals.
These qubits are chosen at regular spacial intervals so that no
information about the structure of the computation is revealed.
It should be noted that in \textbf{Protocol~\ref{prot:UBQC-authentication}}, we allow Alice to reveal
to Bob whether or not she accepts the final result.

Our protocol can also be used in the scenario of non-malicious faults:  because it already uses a fault-tolerant construction,
the measurement of trap qubits in \textbf{Protocol~\ref{prot:UBQC-authentication}} allows for the estimation of the error rate
(whether caused by the environment or by an adversary);  if this error rate is below a certain threshold (this threshold is chosen below the fault-tolerance threshold to take into account sampling errors), Alice accepts the computation.  As long as this is below the
 fault-tolerance threshold, an adversary would still have to guess which qubits
 are part of the code, and which are traps, so Theorem~\ref{thm:authenticated-security} also holds in the fault-tolerant version. The
 only difference is that the adversary can set off a few traps without being detected, but he must still
be able to correctly guess which qubits are in the encoded qubit and which are traps.
  Increasing the security parameters will make up for the fact that Bob can set off a few traps without making the protocol abort.
 This yields a linear trade-off between the error rate and the security parameter. Note that the brickwork state (Figure~\ref{fig:cluster}) can be extended to multiple dimensions, which may be useful for obtaining better fault-tolerance thresholds~\cite{Got00}. While the quantum Singleton bound~\cite{knill-2000-84} allows error
correction codes for which $d_\mathbb{C} \propto n_\mathbb{C}$, it
may be more convenient to use the Toric Code~\cite{Kit97b} for which
$d_\mathbb{C} \propto \sqrt{n_\mathbb{C}}$, as this represents a rather simple encoding while retaining a high ratio of $d_\mathbb{C}$ to $n_\mathbb{C}$.
For the special case of deterministic classical output, a classical repetition code is sufficient and preferable as such an encoding maximizes $n_\mathbb{C}$.
\begin{theorem}[Fault Tolerance]\label{thm:fault-tolerance}
\textbf{Protocol~\ref{prot:UBQC-authentication}} is fault-tolerant.
\end{theorem}
\begin{proof}
By construction,  the circuit created in step~2.1 is fault-tolerant.
Furthermore, the permutation of the circuit wires and insertion of trap qubits  (steps~2.2  and~2.3) preserves the fault tolerance. This is due to the fact that qubits are permuted only within blocks of constant size. The fault-tolerant circuit given in step~2.1 can be written as a sequence of local gates and \textsc{ctrl}-$X$ gates between neighbours. Clearly permutation does not affect the fidelity of local operations. As qubits which are neighbours in the initial fault-tolerant circuit become separated by less than twice the number of qubits in a single block, the maximum number of nearest-neighbour \textsc{ctrl}-$X$ gates required to implement \textsc{ctrl}-$X$ from the original circuit is in~$O(n_\mathbb{C} + 3n_T)$ (the size of a block).
(If required, the multi-dimensional analogue of the two-dimensional brickwork state can be used in order to substantially reduce this distance.) As this upper bound is constant for a given implementation, a lower bound for the fault-tolerance threshold can be obtained simply be scaling the threshold such that the error rate for this worst-case \textsc{ctrl}-$X$ is never more than the threshold for the original circuit. Thus, while the threshold is reduced, it remains non-zero.

Step~\ref{step:convert-to-MBQC} converts the fault-tolerant circuit to a measurement pattern; it is known that this transformation retains the fault-tolerance property~\cite{ND05, AL06}. Finally, in step~\ref{auth:Blind-QC}, distributing the fault-tolerant measurement pattern between Alice and Bob does not disturb the fault tolerance since the communication between them is only classical.
\end{proof}
\begin{theorem}[Blindness]\label{thm:blindness-authenticated-protocol}
\textbf{Protocol~\ref{prot:UBQC-authentication}} is blind while leaking at most $(n,m)$.
\end{theorem}
\begin{proof}
\textbf{Protocol~\ref{prot:UBQC-authentication}} differs from \textbf{Protocol~\ref{prot:UBQC}} in the following two ways: Alice instructs Bob to perform  regular $Z$-basis measurements and she reveals whether or not she accepts or rejects the computation.  It is known that $Z$ measurements change the underlying graph state into a new graph state~\cite{HEB04}. The $Z$ measurements in the protocol are inserted at regular intervals and their numbers are also independent of the underlying circuit computation. Therefore their action  transforms the generic brickwork state into another generic resource still independent of Alice's input and the blindness property is obtained via the same proof of Theorem~\ref{thm:privacy}. Finally, from Alice's decision to accept or reject, only information relating to the trap qubits is revealed to Bob, since Alice rejects if and only if the estimated error rate is too high. The trap qubits are uncorrelated with the underlying computation (in the circuit picture, they do not interact with the rest of the circuit) and hence they reveal no information about Alice's input.
\end{proof}

In the following theorem, for simplicity, we consider the scenario with zero error rate; a proof for the full fault-tolerant version is similar.

\begin{theorem}[Authentication] \label{thm:authenticated-security} For the zero-error case of \textbf{Protocol~\ref{prot:UBQC-authentication}}, if Bob interferes with an authenticated computation, then either he
is detected except with exponentially  small probability (in the security parameter), or his
actions fail to alter the computation.
\end{theorem}
\begin{proof}
If Bob interferes with the computation, then in order for his
actions to affect the outcome of the computation without being detected, he must perform a
non-trivial operation (\textit{i.e.}~an operation other than the identity) on the subspace in which the logical qubits
are encoded. Due to the fault-tolerant construction of Alice's
computation (Theorem~\ref{thm:fault-tolerance}), Bob's operation must have weight at least~$d_\mathbb{C}$. Due to discretization of errors, we can treat Bob's action
as introducing a Pauli error with some probability~$p$.

If a Pauli error acts non-trivially on a trap qubit then the
probability of this going undetected is~${1}/{3}$. Pauli operators
which remain within the code space must act on at least
$d_\mathbb{C}$ qubits. As Bob has no knowledge about the roles of
qubits (Theorem~\ref{thm:blindness-authenticated-protocol}), the probability of him acting on any qubit is equal. As the
probability of acting on a trap is $3 n_T/(n_\mathbb{C} + 3n_T)$, for each qubit upon which he acts non-trivially, the
probability of Bob being detected is $2 n_T/(n_\mathbb{C} +3 n_T)$. Thus the probability of an $M$-qubit Pauli operator going
undetected is below $\left(1-2 n_T/(n_\mathbb{C} + 3n_T)\right)^M$. Since $n_T = n_\mathbb{C}$,
the minimum probability of Bob affecting the computation and going
undetected is $\epsilon = 2^{-d_\mathbb{C}}$.
\end{proof}

\section{Entangled Servers}
\label{sec:entangled-servers}

As stated before, one can view our protocol as an interactive proof system where Alice acts as the verifier and Bob as the prover. An important open problem is to find an interactive proof for any problem in  $\BQP$ with a $\BQP$ prover, but with a purely classical verifier. Our \textbf{Protocol~\ref{prot:UBQC-authentication}} makes progress towards finding a solution by providing an interactive proof for any language in $\BQP$, with a quantum prover and a $\BPP$ verifier that also has the power to generate random qubits chosen from a fixed set and send them to the prover. This perspective was first proposed by Aharonov, Ben-Or and Eban~\cite{Dorit}, however their scheme demands a more powerful verifier.

We present in \textbf{Protocol~\ref{prot:EntBQC}} a solution to another closely related problem, namely the case of a purely classical verifier interacting with two non-communicating entangled provers.
The idea is to adapt \textbf{Protocol~\ref{prot:UBQC}} so that one prover (that we now call a server) is used to prepare the random qubits that would have been generated by Alice in the original protocol, while the other server is used for universal blind quantum computation.
Using the authenticated  protocol (\textbf{Protocol~\ref{prot:UBQC-authentication}}) between Alice and the second server, Alice will detect any cheating servers as clearly, any cheating by Server 1 is equivalent to a deviation from the protocol by Server 2, which is detected in step~\ref{step:Alice-comp} of the protocol, (the proof is directly obtained from Theorem~\ref{thm:authenticated-security}). On the other hand, since Server~2 has access to only half of each entangled state, from his point of view, his sub-system remains in a completely mixed state independently of Server~1's actions and the blindness of the protocol is obtained directly from Theorem~\ref{thm:blindness-authenticated-protocol}.

\begin{algorithm}
\caption{Universal Blind Quantum Computation with Entangled Servers}
 \label{prot:EntBQC}
Initially, Servers~1 and~2 share $\ket{\Phi^+_{x,y}}= \frac{1}{\sqrt{2}}(\ket{00} + \ket{11})$ ($x = 1, \ldots ,n , y = 1, \ldots , m$).

\begin{enumerate}
\item  \textbf{Alice's preparation with Server 1} \\
For each column $x = 1, \ldots , n$ \\
\hspace*{\parindent}\hspace*{\parindent} For each row $y = 1,
\ldots , m$
\begin{enumerate}
\item Alice chooses $$\tilde{\theta}_{x,y} \in_R \{0, \pi/4, 2\pi/4, \ldots, 7\pi/4\}$$ and sends it to Server~1, who measures his part of
$\ket{\Phi^+_{x,y}}$ in   $|\pm_{\tilde{\theta}_{x,y}}\rangle$.
\item Server 1 sends  $m_{x,y}$, the outcome of his measurement,  to Alice.
\end{enumerate}

\item  \label{step:Alice-comp} \textbf{Alice's computation with Server 2} %
\begin{enumerate}
\item Alice runs the authenticated blind quantum computing protocol (\textbf{Protocol~\ref{prot:UBQC-authentication}}) with Server~2,
taking $\theta_{x,y} = \tilde{\theta}_{x,y} + m_{x,y} \pi$.
\end{enumerate}
\end{enumerate}

\end{algorithm}

\section*{Acknowledgements}
We are grateful to Serge Fehr for his help on  Definition~\ref{defn:blind} and Theorem~\ref{thm:privacy}. We would also like to thank Pablo Arrighi,  Fr\'ed\'eric Dupuis,
S\'ebastien Gambs, Matthew McKague, Michele Mosca, Oded Regev, Louis Salvail, Christian Schaffner,  Douglas Stebila and Alain Tapp for
helpful discussions. This work is partially supported by Singapore's National Research Foundation and Ministry of Education and by the  Natural Sciences and Engineering Research Council of Canada.

\newcommand{\SortNoop}[1]{}

\appendix

\section{Measurement-based quantum computing}
\label{appendix:measurement calculus}

We give a brief introduction to the MBQC (a more detailed description is available in~\cite{Jozsa05,Nielsen05,BB06,Mcal07}). Our notation follows that
of~\cite{Mcal07}.

Computations in the MBQC involve the following commands, which are applied to  a single qubit~$i$, or to two qubits~$i$ and~$j$:
\begin{itemize}
\item  Single-qubit
preparations in the state
$\ket{+} = \ost(\ket0 +  \ket1)$\,.

\item Two-qubit entanglement operators  $\et
ij:=\text{\textsc{ctrl}-}Z_{i,j}$.

\item Single-qubit destructive measurements
$\M\alpha i$ defined by orthogonal projections onto $\ket{+_\alpha}=$ {$\ost(\ket0 +  e^{i\alpha}\ket1)$}
(with classical outcome $s_i=0$) and $\ket{-_\alpha} =  \ost(\ket0 -  e^{i\alpha}\ket1)$
(with classical outcome $s_i=1$). Measurement outcomes are summed
(modulo~$2$) resulting in expressions of the form $s=\sum_{i\in I}
s_i$ which are called \emph{signals}.

\item Single-qubit corrections: $\Cx i$,  $\Cz i$
and phase rotations $\Cp i \alpha := e^{\frac{i \alpha Z_i}{2}}$.
Corrections may be dependent on signals, denoted as $\cx i{s}$, $\cz i{s}$ and $\cp i s \alpha$.
\end{itemize}

Dependent corrections on a qubit can be always absorbed in the measurement angle of that qubit:
\begin{align*}
M^\alpha_i \cx i s =& \m{(-1)^s\alpha} i\\
M^\alpha_i \cz i s =& \m{\alpha+s\pi} i\\
M^\alpha_i \cp i s \beta =& \m{\alpha-s\beta} i
\end{align*}

A measurement pattern is described with a finite sequence of commands acting on a finite set of qubits, for which a subset are inputs and a subset are outputs. Measurement patterns are universal for quantum computing.   Patterns are executed from right to left. Any pattern can be rewritten in a \emph{standard} form where all the preparation and entangling command are performed only at the beginning of the computation. This is due to the following commutation relations:
\begin{align*}
\et ij\,\cx is=& \cx is\cz js\,\et ij \\
\et ij\,\cz is=&\cz is\,\et ij\\
\et ij\,\cp is\alpha=&\cp is\alpha\,\et ij
\end{align*}

\section{Universality of the Brickwork state}
\label{appendix:brickwork}
\label{appendix:universal}

\begin{theorem}[Universality]
 The \graphstate $\mathcal{G}_{n \times m}$ is
universal for quantum computation. Furthermore, we only require
single-qubit measurements under the angles $\{0, \pm \pi/4,  \pm \pi/2\}$,
and measurements can be done layer-by-layer.
\end{theorem}

\begin{figure} \centering
   \includegraphics{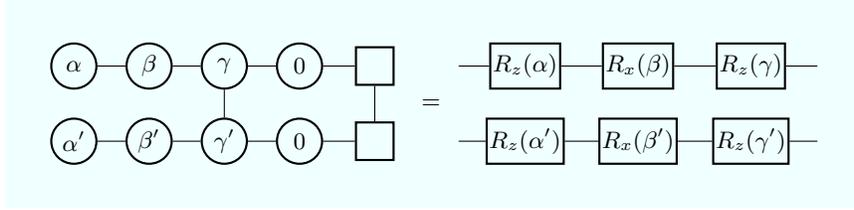}
\caption{Pattern with arbitrary rotations. Squares indicate output qubits.}
    \label{Fig:GeneralRotation}
\end{figure}

\begin{figure} \centering
   \includegraphics{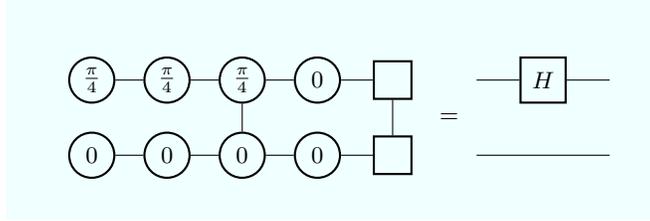}
 \caption{Implementation of a Hadamard gate.}
  \label{Fig:Hadamard}
 \end{figure}

\begin{figure} \centering
   \includegraphics{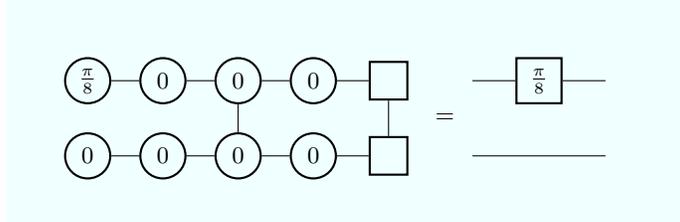}
 \caption{Implementation of a $\pi$/8 gate.}
 \label{Fig:pi/8}
 \end{figure}

\begin{figure} \centering
   \includegraphics{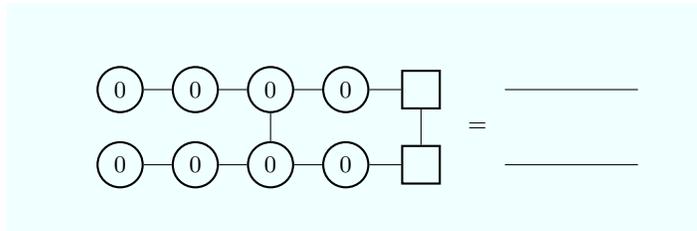}
 \caption{Implementation of the identity.}
 \label{Fig:identity}
 \end{figure}

\begin{figure} \centering
   \includegraphics{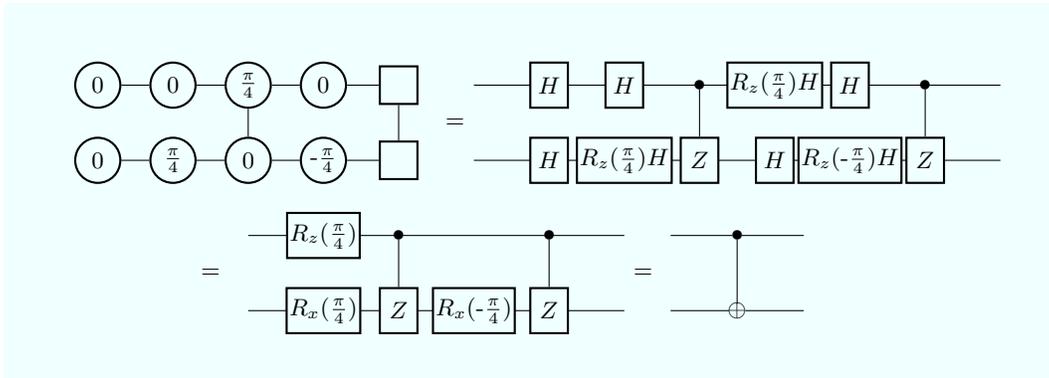}
 \caption{Implementation of a \textsc{ctrl}-$X$.
 }
 \label{Fig:control-Not}
 \end{figure}

\begin{figure}  \centering
\includegraphics{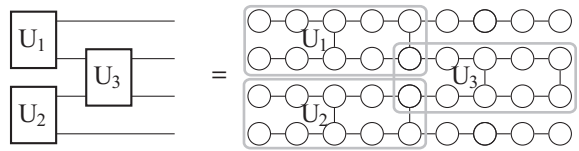}
\caption{Tiling for a 4-qubit circuit with three gates.}
\label{Fig:tiling}
\end{figure}

\begin{proof}
It is well-known that the set $U = \{ \textsc{ctrl-}X, H,
\frac{\pi}{8}\}$ is a universal set of gates; we will show how the brickwork state can be used to compute any gate
in $U$. Recall the \emph{rotation} transformations:  $X(\theta)=e^{\frac{i\theta X}{2}}$ and  $Z(\theta) =  e^{\frac{i\theta Z}{2}}$.

Consider the measurement pattern and underlying graph state given in
Figure~\ref{Fig:GeneralRotation}.  The implicit required corrections
are implemented according to the flow condition~\cite{Flow06} which
guarantees determinism, and allows measurements to be performed
layer-by-layer. The action of the measurement of the first three
qubits on each wire is clearly given by the  rotations  in the
right-hand part of Figure~\ref{Fig:GeneralRotation}~\cite{BB06}. The
circuit identity follows since \textsc{ctrl}-$Z$ commutes with
$Z(\alpha)$ and is self-inverse.

By assigning specific values to the angles, we get the Hadamard gate
(Figure~\ref{Fig:Hadamard}), the  $\pi$/8 gate
(Figure~\ref{Fig:pi/8}) and the identity
(Figure~\ref{Fig:identity}). By symmetry, we can get $H$ or  $\pi$/8
acting  on  logical qubit~2 instead of logical  qubit~1.

In Figure~\ref{Fig:control-Not}, we give a pattern and show using
circuit identities that it implements a $\textsc{ctrl}$-$X$. The
verification of the circuit identities is straightforward. Again by
symmetry, we can reverse the control and target qubits. Note that as
long as we have \textsc{ctrl}-$X$s between any pair of neighbours,
this is sufficient to implement \textsc{ctrl}-$X$ between further
away qubits.

We now show how we can tile the patterns as given in
Figures~\ref{Fig:GeneralRotation} through~\ref{Fig:control-Not} (the
underlying graph states are the same) to implement any circuit
using~$U$ as a universal set of gates. In Figure~\ref{Fig:tiling},
we show how a 4-qubit circuit with three gates, $U_1$, $U_2$ and
$U_3$ (each gate acting on a maximum of two adjacent qubits) can be
implemented on the brickwork state~$G_{9,4}$. We have completed the
top and bottom logical wires with a pattern that implements the
identity.  Generalizing this technique, we get the family of
brickwork states as given in Figure~\ref{fig:cluster} and Definition~\ref{defn:brick}.
\end{proof}


\end{document}